\documentclass[pra,superscriptaddress,twocolumn,showpacs]{revtex4}
\usepackage{amssymb,amsmath,bm}
\usepackage{color}
\usepackage{dsfont}
\usepackage{hyperref}
\usepackage{graphicx}

\definecolor{darkblue}{rgb}{0,0,0.5}
\definecolor{darkred}{rgb}{0.5,0,0}

\hypersetup{
  pdfborder={0,0,0},
  colorlinks=true,
  linkcolor=darkblue,
  citecolor=blue
}

\begin{document}

\title{The Moyal Equation for open quantum systems}

\author{Karl-Peter Marzlin}
\affiliation{Department of Physics, St. Francis Xavier University,
  Antigonish, Nova Scotia, B2G 2W5, Canada}

\author{Stephen Deering}
\affiliation{Department of Physics, St. Francis Xavier University,
  Antigonish, Nova Scotia, B2G 2W5, Canada}

\bigskip

\begin{abstract}
We generalize the Moyal equation, which describes the dynamics
of quantum observables in phase space, to quantum systems
coupled to a reservoir. It is shown that
phase space observables become functionals of fluctuating
noise forces introduced by the coupling to the reservoir.
For Markovian reservoirs, the Moyal equation turns into
a functional differential equation in which the reservoir's
effect can be described by a single parameter.
\end{abstract}



\maketitle

\section{Introduction}
In the study of quantum systems, researchers employ
different methods to achieve their goals. For instance,
the dynamics of a system can be described using the
Schr\"odinger equation for a state or the Heisenberg
equation for operators. Another example is to use the
Wigner function $W(q,p)$ to describe the state by a function
in phase space rather than by a wavefunction $\psi(q)$
or a density matrix $\hat{\rho}$.

The Wigner function \cite{PhysRev.40.749,SchleichPhaseSpace}, which for a one-dimensional system
takes the form 
\begin{align} 
  W(q,p) &= \frac{ 1}{2\pi\hbar} S[\hat{\rho}](q,p)
\label{eq:WignerDef}\\ 
  S[\hat{\rho}](q,p) &=
    \int dq'\,
  \left \langle  q + {\scriptstyle \frac{ 1}{2}} q' \right |
  \hat{\rho} \left | q-{\scriptstyle \frac{ 1}{2}} q' \right \rangle   
   \, e^{-i q' p/\hbar} ,
\label{eq:WeylDef}\end{align} 
has become a very popular tool 
to visualize the state of a quantum system and to
compare it to classical systems. To evaluate mean 
values of observables $\hat{A}$ in phase space one needs to
introduce Weyl symbols $S[\hat{A}](q,p)$, which represent
operators on Hilbert space by functions on phase
space \cite{Weyl1927,case:937}. To keep the notation concise we will
write $A(q,p)$ instead $S[\hat{A}](q,p)$, i.e., the symbol has the same
notation as the operator but without a hat.

The dynamics of the Wigner function is described by
$  \partial_t W = \{H , W \}_M $, where 
$H$ is the Weyl symbol of the Hamiltonian. The
Moyal bracket of two Weyl symbols $A, B$ is defined as
\cite{PSP:1593184,Groenewold1946405,Osborn199579,deAlmeida1998265}
\begin{align} 
  \{A,B\}_M &=  \frac{ 1}{i\hbar} (A\star B - B\star A).
\end{align} 
The star product $\star$ between two symbols is a representation
of the non-commutative product between operators in 
phase space \cite{STR57,FAB72,Gro76,Roy77,jpa42-415302}.
For our purpose it can be written as
\begin{align} 
  A(q,p)\star B(q,p) &= 
  A({\cal L})B(q,p) \;  =\; 
  B({\cal R})A(q,p)
\\
   A({\cal L}) &= A\left ( q+i \frac{ \hbar}{2} \partial_p\; ,\;
     p-i \frac{ \hbar}{2} \partial_q  \right )
\label{eq:starLeft}\\
  B({\cal R}) &= B\left ( q-i \frac{ \hbar}{2} \partial_p\; ,\;
     p+i \frac{ \hbar}{2} \partial_q  \right ).
\label{eq:starRight}\end{align} 
In the limit $\hbar \rightarrow 0$, the Moyal bracket turns into
the classical Poisson bracket and the dynamics of the Wigner
function is described by the Liouville equation \cite{PSP:1593184}.

The dynamical equation for the Wigner function
is the analogue of the Schr\"odinger equation in phase space.
Likewise, the analogue of the Heisenberg equation of motion for operators
$\hat{A}$ is the Moyal equation
\begin{align} 
   \partial_t  A &= \{A ,  H  \}_M .
\end{align} 
By combining the two possibilities (Hilbert space and phase space)
to represent a quantum system with the two possibilities
to describe its dynamics (Schr\"odinger picture and Heisenberg
picture), we thus arrive at four different ways to study its evolution.

The dynamical equations above apply to perfectly isolated quantum
systems. However, for a realistic description of experiments,
the coupling to the environment has to be taken into account.
This is accomplished by considering open 
quantum systems \cite{GardinerZollerQuantumNoise,BreuerPettrucione},
where the system of interest is coupled to another system
(the reservoir) that cannot be accessed in an experiment. 
Loss of information about correlations with the reservoir will 
introduce decoherence and noise to the system. In a similar way,
the measurement process can be generalized by coupling
the system to a second system that represents 
the detector \cite{BuschLahtiMitt}.

In open quantum systems, the four dynamical equations discussed above 
are replaced by more general equations. 
The Schr\"odinger equation is replaced by the 
well-known master equation
\cite{linblad:cmp76}, while the Heisenberg equation of motion
for operators turns into the quantum Langevin 
equation \cite{GardinerZollerQuantumNoise,BreuerPettrucione},
which typically takes the form
\begin{align} 
  \partial_t \hat{A} &= -\frac{ i}{\hbar} [\hat{A},\hat{H}]  
  -\frac{ i}{2\hbar} \{ [\hat{B},\hat{A}], \hat{F}
  -\gamma \partial_t\hat{B} \},
\label{eq:QLangevin}\end{align} 
where $\{ ., .\}$ denotes the anti-commutator, $\hat{B}$ is the system
operator that is involved in the coupling to the reservoir, 
$\hat{F}$ is a noise operator, and $\gamma$ a decoherence rate.

In phase space, the evolution of the Wigner function for open quantum
systems leads generally to a strictly positive 
Wigner function \cite{0305-4470-35-11-312,PhysRevE.69.016204},
thus introducing classicality \cite{RevModPhys.75.715}.
Its dynamical equation  takes the
form of a Fokker-Planck equation \cite{PhysRevA.4.739,0022-3700-18-8-019}.

To our knowledge, the extension of the Moyal equation, i.e., the
last of the four dynamical equations, 
has only been addressed in
general terms by Ozorio de Almeida \cite{quant-ph/0612029}.
In this paper we aim to shed more light on this case by studying the Moyal
equation for a Schr\"odinger 
particle coupled to a thermal reservoir of harmonic oscillators.
Our main result is that for Markovian
reservoirs, in which correlations decay on a short time scale,
the Moyal equation takes the form
\begin{align} 
   \partial_t  A(q,p,t) &= \{A,  H  \}_M 
   -2\gamma p \frac{ \delta A}{\delta F(t)}
   +F(t) \partial_p  A,
\label{eq:MoyalLangevin3}\end{align} 
where $F(t)$ is a fluctuating force introduced by the reservoir, and
$\delta A /\delta F(t)$ denotes the functional derivative
with respect to this force. The decoherence rate $\gamma$
determines the time scale $\gamma^{-1}$ on which information
about reservoir correlations is lost.

\section{Open Moyal equation}
To derive an example for an open Moyal equation, we consider
the model proposed by Ford, Kac 
and Mazur \cite{FordKacMazur,FordKac1987}, in which a single,
one-dimensional Schr\"odinger particle is coupled to 
a reservoir of $N$ harmonic oscillators. 
The Weyl symbol of the Hamiltonian is given 
by \footnote{The Hamiltonian operator takes the same form, with the
symbols $q,p,q_n,p_n$ replaced by the respective operators
because $S[g(\hat{q})]=g(q)$ for any operator
$g(\hat{q}) $ that is a function of $q$ only. A similar
statement holds for operators of the form $g(\hat{p}) $.}
\begin{align} 
  H(q,p,q_n,p_n) &= H_S+H_\text{int}+H_R
\\
  H_S &= \frac{ 1}{2m}p^2 +V(q)
\\ H_R+H_\text{int}&=
      \sum_{n=1}^N \left (
  \frac{ 1}{2m_n} p_n^{2}+ \frac{ 1}{2}k_n (q_n-q)^{2} \right ),
\label{eq:WeylRint}\end{align} 
with $k_n = m_n \omega_n^2$.
Here $q,p$ are position and momentum of the system particle,
and $q_n , p_n$ are the respective quantities for the $n$th
reservoir oscillator. The symbol of $H_R$ of the reservoir
Hamiltonian alone can be found
by setting $q=0$ in Eq.~(\ref{eq:WeylRint}).
Using the two different representations
(\ref{eq:starLeft}), (\ref{eq:starRight}) of the star product, the
Moyal bracket can be written as
\begin{align} 
    \{A,H\}_M &= 
   \frac{ 1}{i\hbar}  \big(H({\cal R})-H({\cal L}) \big ) A(q,p,q_n,p_n,t).
\label{eq:M1}
\end{align} 
It therefore corresponds to linear differential
operator on phase space that is defined through Eq.~(\ref{eq:M1}).
It is shown in App.~\ref{sec:AppMoyalDerivation} that
by a transformation
\begin{align} 
  A(t)&= e^{t \{. , H_R \}_M } e^{q K(t)}\bar{A}(t),
\label{eq:AAbarTrafo}
\end{align} 
the Moyal equation can be cast into the form
\begin{align} 
  \partial_t\bar{A} &=
   \{ \bar{A},H_S\}_M + \frac{ p}{m} K(t) \bar{A}  +  F(t) 
   \, \partial_p \bar{A} .
\label{eq:MoyalLangevin2}
\end{align} 
Here $F(t)$ is a function of the reservoir's phase space variables
$q_n,p_n$ and $K(t)$ is a differential operator acting on 
functions of the reservoir variables.
Their definition is given in Eqs.~(\ref{eq:Fdef})
and (\ref{eq:Kdef}), but for practical purposes only the relation
\begin{align} 
   K(t) F(t') &= - C(t-t')
\label{eq:KFaction}\\
  C(t) &= \sum_{n=1}^N k_n
  \cos(\omega_n t)
\label{eq:defC}
\end{align} 
is relevant. The term proportional
to $F(t) $ in Eq.~(\ref{eq:MoyalLangevin2})
could be introduced by modifying
the system's potential energy as
\begin{align} 
  V(q) &\rightarrow V(q) - q F(t) .
\end{align} 
Therefore, $F(t) $ can be interpreted as a time-dependent
homogeneous force acting on the system. Because it
depends on the reservoir variables, it will fluctuate with the state
of the reservoir. We therefore can interpret $F(t) $
as the phase-space symbol of
a fluctuating random force. It is the equivalent of
noise operators that appear in the quantum Langevin equation
(\ref{eq:QLangevin}) and therefore may be called a noise (Weyl) symbol.

As $K(t)$ corresponds to a derivative operator, it does not
represent a phase space symbol, but rather a phase-space
super-operator that maps the symbol of an  operator to another symbol.
We will see below that $K(t)$ describes the dissipation 
in the system associated with the fluctuating random force.

The function $C(t)$ of Eq.~(\ref{eq:defC}) can be interpreted
as a force correlation function. It is shown in 
App.~\ref{sec:forceCorrelation} that for a thermal
reservoir of temperature $T \gg \hbar\omega_n/k_B$ one has
\begin{align} 
  \langle F(t) F(t')\rangle &= 
 k_B T \, C(t-t').
\label{eq:FFcorrelation}\end{align} 

\section{Markovian Moyal equation}
In many cases of interest, the correlation function (\ref{eq:defC})
decays on a time scale $\tau$  that is much shorter than the
time scales relevant for the evolution of the system. 
A physical example would be an atom
as a system that is coupled to the quantized electromagnetic field
as reservoir. This coupling is responsible for spontaneous decay
of an excited atom, which for optical transitions happens 
on a time scale of nano seconds. The electromagnetic correlation function
at these frequencies decays on the scale of femto seconds.

When the reservoir correlation function decays quickly, it is possible
to make the Markov approximation. To analyze the Markov approximation
in phase space, we first rewrite Eq.~(\ref{eq:MoyalLangevin2}) as an
integral equation,
\begin{align} 
  \bar{A}(t)&= e^{t \{ .,H_S \}_M}\bar{A}(0)
  + \int_0^t dt' e^{(t-t') \{ .,H_S \}_M}
\nonumber \\ &\hspace{4mm}\times
  \left (\frac{ p}{m} K(t') +  F(t') 
   \, \partial_p  \right ) \bar{A}(t') .
\label{eq:DysonInt}\end{align} 
We note that $\bar{A}(t,q,p,q_n,p_n)$ is a function of both
system and reservoir variables, while for a system operator
the initial Weyl symbol $\bar{A}(0,q,p)$ does not depend
on the reservoir. By expanding Eq.~(\ref{eq:DysonInt}) into
a Dyson series one can see that $ \bar{A}(t)$ depends on $F(t')$
through integrals of the form
\begin{align} 
   I=\int_0^tdt' g(t') F(t'),
\end{align} 
with $g(t')$ being known functions, 
and convolutions of such integrals. The action of $K(t)$ on such
expressions  amounts to
\begin{align} 
  K(t) I &= -\int_0^t dt' g(t') C(t-t').
\end{align} 
To perform the Markov approximation, we define the decay rate
\begin{align} 
  \gamma &= \frac{ 1}{m} \int_0^\infty dt\,  C(t).
\end{align} 
Then, for times $t\gg \tau$, we find
\begin{align} 
  K(t) I &= -\int_0^t dt'' g(t-t'') C(t'')
\\ &\approx -g(t) \int_0^t dt'' C(t'')
\\ &= -m \gamma g(t)
\\ &= -2m\gamma \frac{ \delta I}{\delta F(t)}. 
\label{eq:MarkovApprox}
\end{align} 
With this approximation
\footnote{While $\delta I/\delta F(t') = g(t')$ for all points
  $0<t'<t$, one has to be careful at the boundary of the integral $I$.
The factor of 2 in Eq.~(\ref{eq:MarkovApprox}) is a consequence of the
following procedure. We rewrite $I$ by replacing the integration boundary $t$ by
$\infty$ and the function $g(t')$ by $g(t')\theta(t-t')$, with $\theta$
the step function. Then $\delta I/\delta F(t) =
g(t)\theta(0)$. Setting $\theta(0)=1/2$ yields
Eq.~(\ref{eq:MarkovApprox}).}, 
the open Moyal equation 
takes the form (\ref{eq:MoyalLangevin3}).
In the following section we will verify this approximation
at the example of a free particle.

\section{Free particle coupled to a reservoir}\label{Sec:freeParticle}
To illustrate the general framework of the open Moyal equation
we consider the situation in which the system particle is not
subject to an external potential, $V=0$. In this case
the Markovian Moyal equation takes the simple form
\begin{align} 
   \partial_t  \bar{A} &= 
  \frac{ p}{m} \partial_q \bar{A}
   -2\gamma p \frac{ \delta \bar{A}}{\delta F(t)}
   +F(t) \partial_p  \bar{A} .
\label{eq:freeMoyal}\end{align} 
\subsection{Canonical Weyl symbols}\label{sec:canonicalSymbols}
We first solve Eq.~(\ref{eq:freeMoyal}) for the canonical variables, 
where $A(0)=q$ or $A(0)=p$,
by making the ansatz
\begin{align} 
  \bar{A}(q,p,t) &= \beta_1(t) p+\beta_2(t)q+\bm{ \beta}_3(t),
\end{align} 
where $\beta_1, \beta_2$ are functions of time only
while $\bm{\beta}_3(t)=\bm{\beta}_3(t,q_n,p_n)$ 
may also depend on
the phase-space variables of the reservoir.
We use bold greek letters to indicate such a dependence.
Inserting this into Eq.~(\ref{eq:freeMoyal}) and sorting the
result with respect to $q$ and $p$ yields
\begin{align} 
  \partial_t \beta_1 &= \frac{ 1}{m} \beta_2  -2 \gamma
  \frac{ \delta \bm{\beta}_3}{\delta F(t)}
\label{eq:beta1Dgl} \\
   \partial_t \beta_2 &=0
\\
  \partial_t\bm{\beta}_3&= F(t) \beta_1 ,
\end{align} 
so that $\beta_2(t) = \beta_2(0)$ and
\begin{align} 
  \bm{\beta}_3(t) &= \int_0^tdt'\, 
   \beta_1(t') F(t').
\label{eq:A3sol} \end{align} 
Inserting this into Eq.~(\ref{eq:beta1Dgl}) yields
\begin{align} 
  \partial_t \beta_1 &= \frac{ 1}{m} \beta_2  - \gamma
  \beta_1,
\label{eq:canonBeta1Dgl}\end{align} 
which is solved by
\begin{align} 
  \beta_1(t) &= \beta_1(0) e^{-\gamma t} +\frac{
    \beta_2(0)}{m\gamma}
  (1-e^{-\gamma t}).
\label{eq:solA1} \end{align} 
For the symbols of position (momentum) we have 
$\beta_1(0)=0$ and $\beta_2(0)=1$
($\beta_1(0)=1$ and $\beta_2(0)=0$), respectively, so that
\begin{align} 
  \bar{A}_q(t) &= 
   q+ \frac{p}{m\gamma}
  ( 1-e^{-\gamma t}) +
      \frac{1}{m\gamma}
   \int_0^t dt' \,
  (1-e^{-\gamma t'}) F(t')
\\
  \bar{A}_p(t) &= 
    e^{-\gamma t}p  +
   \int_0^t dt' \,  e^{-\gamma t'}F(t').
\end{align} 
Here we have adopted the notation that an 
index at the symbol of an operator refers to
its initial value, e.g., 
$\bar{A}_q(0) = q$.
It remains to perform transformation (\ref{eq:AAbarTrafo}).
By using $e^{L_R t} F(t') = F(t'-t)$ and
approximation (\ref{eq:MarkovApprox}) we obtain
\begin{align} 
  A_q(t) &= 
   q e^{-\gamma t}+ \frac{p}{m \gamma}
  (1-e^{-\gamma t})
\nonumber \\ &\hspace{4mm}
  +\frac{1}{m \gamma}
   \int_0^t dt' \,
  ( 1-e^{-\gamma t'}) F(t'-t)
\label{eq:symq}\\
  A_p(t) &= 
    e^{-\gamma t} (p-m\gamma q)  +
   \int_0^t dt' \,  e^{-\gamma t'}F(t'-t).
\label{eq:symp}\end{align} 
We have verified that this solution agrees with the corresponding
operator-valued solution of the quantum Langevin equation.
Furthermore, it is shown in App.~\ref{sec:LaplaceTreatment} 
that the Markovian solution presented here agrees with an exact
treatment based on a specific correlation function $C(t)$.

To turn solutions (\ref{eq:symq}) and (\ref{eq:symp})
into symbols for the system particle alone,
we have to take the average with respect to the reservoir.
It is shown in 
App.~\ref{sec:forceCorrelation} that for a thermal
reservoir the mean noise force vanishes, $\langle F(t) \rangle =0$.
Therefore, the open Weyl symbols of position and momentum
take the form  (\ref{eq:symq}) and (\ref{eq:symp}) with 
$F(t)$ set to zero.

The physical interpretation of this motion is as follows. The coupling
to many oscillators with different frequencies results in a dissipation
of energy from the particle into the reservoir. Therefore, the
momentum of the particle is damped on a time scale $\gamma^{-1}$.
The damping of the initial position $q$ arises because in the
model by Ford, Kac and Mazur all oscillators pull the system particle
toward the common equilibrium point at $q=0$. However, the particle
continues to move in its initial direction for a time of the order of
 $\gamma^{-1}$, which explains the term proportional to $p/(m\gamma)$
in $A_q(t)$.

The reader may have noticed that $A_p(0) = p-m\gamma q$ 
does apparently not fulfill the correct initial conditions. However, this is merely
a consequence of the fact that the Markovian approximation is only
valid for times $t\gg \tau$. The non-Markovian derivation in
App.~\ref{sec:LaplaceTreatment} shows that the term proportional
to $m\gamma q$ builds up on the short time scale $\tau$ and then decays
on the long time scale $\gamma^{-1}$.

\subsection{Variance Weyl symbols}\label{sec:qpSqrd}
To fully appreciate the influence of the reservoir one also
needs to study the uncertainty of the canonical variables.
We therefore consider symbols with initial condition
$\bar{A}(0) = q^2$ or $\bar{A}(0)=p^2$ by making the ansatz
\begin{align} 
  \bar{A}&= \beta_1 p^2 +\beta_2 p q +\beta_3 q^2
  + \bm{\beta}_4 p +\bm{\beta}_5 q +\bm{\beta}_6 .
\label{eq:varianceAnsatz}\end{align} 
As before, we sort the terms with respect to the
power of $q$ and $p$. For four of the six coefficients
this can be done exactly as for the canonical symbols, leading to
$\beta_3(t)=\beta_3(0)$ and
\begin{align} 
  \beta_2(t) &= \frac{ 2\beta_3(0)}{m\gamma}(1-e^{-\gamma t})
\\
  \bm{\beta}_5(t) &= \int_0^tdt'\, \beta_2(t') F (t')
\\
  \bm{\beta}_6(t) &= \int_0^tdt'\, \bm{\beta}_4(t') F (t').
\label{eq:beta6}\end{align} 
The remaining two coefficients obey
\begin{align} 
  \partial_t\beta_1 &= \frac{ 1}{m}\left  (\beta_2 -2\gamma
  \frac{ \delta \bm{\beta}_4}{\delta F(t)} \right )
\\
  \partial_t\bm{\beta}_4 &= 2\beta_1 F 
  + \frac{ 1}{m} \left (\bm{\beta}_5
  -2\gamma
  \frac{ \delta \bm{\beta}_6}{\delta F(t)}  \right ).
\end{align} 
Using Eq.~(\ref{eq:beta6}) leads to
\begin{align} 
  \bm{\beta}_4(t) &= 2\int_0^t dt'\, e^{\gamma(t'-t)}
   \beta_1(t') F(t') 
\nonumber \\ &\hspace{4mm}
  +  \frac{ 1}{m}\int_0^t dt'\, \int_0^{t'} dt''\,  e^{\gamma(t'-t)}
   \beta_2(t'') F(t'') .
\label{eq:A4formal}\end{align} 
The functional derivative of $\bm{\beta}_4$ can then be 
simplified by noting that, for an arbitrary function $g(t,t'')$,
\begin{align} 
  \int_0^t dt'\, \int_0^{t'} dt''\,g(t',t'')
   &= \int_0^t dt''\, \int_{t''}^{t} dt'\,g(t',t''),
\end{align} 
to obtain
\begin{align} 
  \partial_t \beta_1 &= \frac{ 1}{m} \beta_2 -2\gamma \beta_1,
\end{align} 
which is solved by
\begin{align} 
  \beta_1(t) &= e^{-2\gamma t}\beta_1(0) +
  \frac{\beta_3(0)}{m^2\gamma^2} (1-e^{-\gamma t})^2.
\end{align} 
Applying transformation (\ref{eq:AAbarTrafo}) to 
Eq.~(\ref{eq:varianceAnsatz}) and using the
expressions for $\bm{\beta}_i \; (i=4,5,6)$ yields
\begin{align} 
  A(t) &= \bar{A}(t) -m \gamma q (
   2( p-m\gamma q) \beta_1 + q \beta_2
 +\bm{\beta}_4 ) ,
\label{eq:BvarSol}\end{align} 
where $F(t')$ is replaced by $F(t-t')$ everywhere.

Because $\langle F(t) \rangle =0$, 
the averaged form of Eq.~(\ref{eq:BvarSol}) takes the form
\begin{align} 
  A(t) &= 
  \beta_1 p^2
   +q^2 (\beta_3 -m\gamma\beta_2
  +2 m^2\gamma^2 \beta_1)
\nonumber \\ &\hspace{4mm}
+ pq (\beta_2   -2m\gamma \beta_1 
  )+\langle \bm{\beta}_6(t) \rangle .
\end{align} 
The mean value of  $\bm{\beta}_6$ 
 can be evaluated using Eqs.~
(\ref{eq:beta6}), (\ref{eq:A4formal}), and
(\ref{eq:FFcorrelation}).
Within the Markovian approximation, the result is given by
\begin{align} 
   \langle \bm{\beta}_6(t) \rangle &=
  2 k_B T m \gamma \int_0^t dt' \beta_1(t')
\\ &=
  m k_B T \Big \{
  \beta _1(0) 
   \left(1-e^{-2 \gamma  t}\right)
\nonumber \\ &\hspace{4mm}
  + \frac{\beta _3(0) \left(2 \gamma  t-e^{-2 \gamma  t}+4
   e^{-\gamma t}-3\right)}{\gamma ^2 m^2}
  \Big \}.
\end{align} 
It remains to apply the initial conditions, $\beta_1(0)=1$ for
$A_{p^2}(t)$ and $\beta_3(0)=1$ for
$A_{q^2}(t)$, with all other coefficients being zero.
Putting everything together we arrive at the 
open Weyl symbols
\begin{align} 
  A_{q^2}(q,p,t) &= A_q^2 + 
 q^2 \left(1- e^{-\gamma t}\right)^2
\nonumber \\ &\hspace{4mm}
 + \frac{ k_B T}{m\gamma^2}(2 \gamma  t-3 - e^{-2 \gamma  t}+4
 e^{-\gamma t})
\\
  A_{p^2}(q,p,t) &=A_p^2
  + m k_B T \left(1- e^{-2 \gamma  t}\right)
  +e^{-2 \gamma  t} m^2  \gamma ^2 q^2.
\end{align} 
We have confirmed that this solution agrees with the
operator-valued solution derived from the quantum Langevin
equation. The physical interpretation is as follows. The term 
$2 k_B T t/(m \gamma)$ in $A_{q^2}$ corresponds to a diffusion
of the particle position. For solutions of the diffusion equation
$\partial_t f(t,q) = D \partial_q^2 f(t,q)$,
the variance of position increases as $2 D t$ for sufficiently large
times. This implies that the coupling to the reservoir can be
linked to a diffusion coefficient $D= k_B T /(m \gamma)$.
The Einstein-Smoluchowski 
relation $D=\mu k_B T$  then implies that the mobility
of the system particle is given my $\mu =1/(m\gamma)$.

The term proportional to $m k_B T$ in $A_{p^2}$ describes
the thermalization of the system particle through its coupling
to the reservoir. Keeping in mind that $\langle E_\text{kin} \rangle 
= \langle A_{p^2} \rangle /(2m)$ one can see that for times
$t\gg \gamma^{-1}$ the kinetic energy of the reservoir approaches
$k_B T/2$, confirming the equipartition theorem.
Finally, the terms proportional to $q^2$ are a consequence of the
dragging towards the origin in the Ford-Kac-Mazur model that we
discussed above.

\section{Discussion and Conclusion}
In the previous sections we derived the open Moyal equation
(\ref{eq:MoyalLangevin3}),
which describes the dynamics of open quantum systems coupled
to a Markovian reservoir, and illustrated its features at the example
of a free particle. The significance of the open Moyal equation
is not so much in the specifics given here, but in the possibility
to extend this model to describe open systems of larger interest.

This is the same situation as with master equation and 
quantum Langevin equation (\ref{eq:QLangevin}). Both equations
have been derived from specific models, but have been
generalized to describe the influence of various reservoir-induced
effects on a large variety of systems. Examples include 
spontaneous emission, thermal excitation, spin dephasing, 
vibrational relaxation in molecules, and lossy
optical cavities \cite{GardinerZollerQuantumNoise,BreuerPettrucione}.

In practice, most researchers do not derive reservoir properties
from first principles but rather pick an ad hoc model to include
the effect of a reservoir. This choice is not arbitrary
but has to obey general principles. Positivity and trace preservation
of the density matrix limit Markovian master equations to the 
celebrated Lindblad form \cite{linblad:cmp76}. Likewise, noise operators
$\hat{F}(t)$ and decoherence rate $\gamma$ in quantum Langevin
equations are related through a fluctuation dissipation theorem.

The same is true for the open Moyal equation. A rather trivial
extension of the free particle discussed above is a particle under the
influence of a constant external force $F_0$. This can be accomplished by
including a linear potential $V(q)=-q F_0$ in Eq.~(\ref{eq:LS}).
Using the techniques of Sec.~\ref{Sec:freeParticle} it is not hard to see
that the only change in solutions (\ref{eq:symq}), (\ref{eq:symp}) is
then to replace $F(t)$ by $F(t)+F_0$. The Weyl symbols of position and
momentum are then modified according to
\begin{align} 
  A_q(t) &\rightarrow A_q(t) + \frac{ F_0}{m\gamma} 
  \left ( t -\gamma^{-1} (1-e^{-\gamma t}) \right )
\\ 
  A_p(t) &\rightarrow A_p(t) + \frac{ F_0}{\gamma} (1-e^{-\gamma t}) .
\end{align} 
For large time $t$ this describes a particle moving with
drift velocity $F_0/(m\gamma)$, which confirms the result for the
mobility $\mu$ found in Sec.~\ref{sec:qpSqrd}. Hence,
the system's response to a linear perturbation is linked to
the diffusion of $q$ through the Einstein-Smoluchowski 
relation, which is one of the earliest examples of
a fluctuation dissipation theorem.

The model presented here can be readily extended in several ways.
An obvious extension is to combine the model by Ford, Kac and Mazur
with more complicated external potentials $V(q)$. An interesting
question arises for nonlinear potentials of Kerr type, where the 
exact solution for Weyl symbols exhibits a singularity
\cite{jpa42-415302}. Because dissipation often has a moderating
effect, it may be possible that coupling to a reservoir may
eliminate this singularity. One may also consider the effect of a
reservoir on a system that consists of two or more interacting
particles, for instance for two photons interacting via
cross Kerr modulation \cite{JOSA-B27-A36}. 

Another extension would be to generalize the quadratic coupling
between the position of system and reservoir particles in
the Ford-Kac-Mazur model. One way would be to introduce
a distribution of equilibrium positions for the reservoir oscillators
and to average over this distribution. This may eliminate
the drag towards the origin that we discussed in
Sec.~\ref{Sec:freeParticle}. One may also introduce a different
coupling like $p\sum_n p_n$ in Eq.~(\ref{eq:WeylRint}), which
could be realized by a series of quantum LC circuits coupled 
through their mutual inductance \cite{Meng20092027}.
In the context of quantum Langevin equations, it is 
possible to extend the interaction with a reservoir to 
nonlinear couplings \cite{PRL73:240,shapiro:062305,JOSA-B27-A36},
although this is considerably more involved.
It is conceivable that the same could be accomplished for
the open Moyal equation. 

Finally one could follow the common practice in the field
of quantum Langevin equations and simply pick an ad hoc model
to obtain another open Moyal equation. This could be done
by replacing $p$ in Eq.~(\ref{eq:MoyalLangevin3}) by another 
system observable $O$, and $F(t)$ by random
fluctuations of the corresponding time derivative $\partial_t O$.
Given that phase
space methods are an excellent tool to compare classical and
quantum dynamics, and given that coupling a quantum system
to a reservoir generally suppresses quantum correlations, 
the open Moyal equation may therefore provide a promising method
to study how classical behaviour emerges in open quantum systems.

\acknowledgments
We are indebted to Dr. Thomas Osborn for detailed comments on the
manuscript and Dr.~Brandon van Zyl for helpful discussions.
K.-P.~M. and S.~D.~wish to thank NSERC for a Discovery Grant and
an Undergraduate Student Research Award, respectively.

\begin{appendix}
\section{Derivation of the non-Markovian open Moyal
  equation}\label{sec:AppMoyalDerivation}
To avoid a cluttered notation, we will define linear operators
$L_i $ through
\begin{align} 
  L_i A &:= \{A ,H_i \}_M  \; , \; i=S,R,\text{int} .
\end{align} 
The explicit form of the Moyal bracket for the various parts of the
Hamiltonian is then given by
\begin{align} 
   L_S  &= \frac{ p}{m} \partial_q  + \frac{ 1}{i\hbar} [
   V (q-i 
   {\scriptstyle\frac{\hbar}{2}} \partial_p)
  -   V\left (q+i  {\scriptstyle\frac{\hbar}{2}} \partial_p\right )
  ] 
\label{eq:LS}\\
  L_R+  L_\text{int} &  = 
   \sum_{n=1}^N \left [ \frac{p_n }{m_n}\partial_{q_n} 
  -k_n (q_n-q) (\partial_{p_n}-\partial_p) \right ].
\end{align} 
To solve the reservoir part of Eq.~(\ref{eq:M1}),
it is useful to switch to complex coordinates 
\begin{align} 
  \alpha_n &= \frac{ 1}{\sqrt{2}} \left (\frac{ q_n}{l_n}+i \frac{
    l_n}{\hbar} p_n \right ) 
\\ l_n &= \sqrt{\frac{ \hbar}{m_n\omega_n}},
\end{align}
which yields
\begin{align} 
    L_\text{int}  &=-q \partial_p \sum_{n=1}^N k_n 
  + i q \sum_{n=1}^N \frac{ k_nl_n}{\sqrt{2}\hbar} 
  (\partial_{\alpha_n}-\partial_{\alpha_n^*})
\nonumber \\ &\hspace{5mm}
    + \partial_p \sum_{n=1}^N \frac{ k_nl_n}{\sqrt{2}} 
    (\alpha_n+ \alpha_n^*) 
\\
   L_R  &= \sum_{n=1}^N  i \omega_n ( \alpha_n^* \partial_{\alpha_n^*} 
   - \alpha_n \partial_{\alpha_n} ).
\end{align} 
We now go into an interaction picture with respect to the reservoir
by setting
\begin{align} 
  A(q,p,\alpha_n, \alpha_n^*) &= e^{t L_R} \tilde{A} (q,p,\alpha_n, \alpha_n^*).
\end{align} 
The Moyal equation then takes the form
\begin{align} 
  \partial _t \tilde{A} &= e^{-t L_R } (L_S+L_\text{int}) e^{t L_R }\tilde{A}
\\ &= L_S \tilde{A}   - q \partial_p \sum_{n=1}^N k_n  \tilde{A} + 
  \sum_{n=1}^N \frac{ k_nl_n}{\sqrt{2}} 
\nonumber \\ & \hspace{5mm}
  \times  \Big  \{ 
   \frac{ i}{\hbar} q e^{-t L_R} \left (\partial_{\alpha_n}-\partial_{\alpha_n^*}
                   \right ) e^{t L_R} 
\nonumber \\ & \hspace{10mm}+ 
    e^{-t L_R}  (\alpha_n+ \alpha_n^*) e^{t L_R}  \partial_p 
  \Big \}\tilde{A}.
\end{align} 
Operators of the form $O(t) = e^{-t L_R} O(0) e^{t L_R} $
obey the differential equation 
$\partial_t O = -[L_R,O]$. It is not hard to see that for
$O(0)=\alpha_n$ and $O(0)=\partial_{\alpha_n}$, this equation
has the solution 
\begin{align} 
     e^{-t L_R}  \alpha_n e^{t L_R}  &= e^{i\omega_n t} \alpha_n
\\
     e^{-t L_R}  \partial _{\alpha_n} e^{t L_R}  &= e^{-i\omega_n
       t} \partial _{\alpha_n} .
\end{align} 
Hence the Moyal equation can be rewritten as
\begin{align} 
  \partial _t \tilde{A} &= 
  \left (L_S - q \partial_p \sum_{n=1}^N k_n  \right ) \tilde{A} +
  \sum_{n=1}^N \frac{ k_nl_n}{\sqrt{2}} 
\nonumber \\ & \hspace{5mm}\times
  \Big  \{\frac{  i}{\hbar} q \left ( e^{-i\omega_n t}\partial_{\alpha_n}- e^{i\omega_n t}\partial_{\alpha_n^*}
                   \right ) 
\nonumber \\ & \hspace{10mm}+ 
   ( e^{i\omega_n t}\alpha_n+  e^{-i\omega_n t}\alpha_n^*) \partial_p 
  \Big \}\tilde{A}
\\ &= \{ L_S  - C(0) q \partial_p +\dot{K} q + F(t) \, \partial_p \}\tilde{A},
\label{eq:MoyalLangevin}\end{align} 
with
\begin{align} 
  F(t) &= \sum_{n=1}^N \frac{ k_nl_n}{\sqrt{2}} 
  \left \{    e^{i\omega_n t}\alpha_n
   +e^{-i\omega_n t}\alpha_n^* \right \}
\label{eq:Fdef}
\\
  K(t) &=  -\sum_{n=1}^N \frac{ k_nl_n}{\sqrt{2}\hbar\omega_n} 
   \left \{   e^{-i\omega_n t} \partial_{\alpha_n}
   +e^{i\omega_n t} \partial_{\alpha_n^*} \right \}
\label{eq:Kdef}\end{align} 

To perform the Markovian approximation it is useful to
reformulate Eq.~(\ref{eq:MoyalLangevin}) by setting
$\tilde{A}(t) = \exp (q K(t)) \bar{A}(t)$. 
Because $[K(t),K(t')]=0$ we can
use the Baker-Campbell-Hausdorff formula
and Eq.~(\ref{eq:KFaction}) to transform
the open Moyal equation into
\begin{align} 
  \partial_t \bar{A} &=
    \{e^{-q K(t)} L_S e^{q K(t)} - C(0) q \partial_p
\nonumber \\ &\hspace{5mm}
  +  e^{-q K(t)}F(t) e^{qK(t)}
   \, \partial_p \} \bar{A}
\\ &= \{ L_S + \frac{ p}{m} K(t)  - C(0) q \partial_p
\nonumber \\ &\hspace{5mm}
  +    (F(t) -q [K(t) ,F(t)])
   \, \partial_p \} \bar{A}
\\ &= \{ L_S + \frac{ p}{m} K(t) + 
   F(t) 
   \, \partial_p \} \bar{A},
\end{align} 
which is Eq.~(\ref{eq:MoyalLangevin2}).

\section{Averaging over the reservoir degrees
of freedom}\label{sec:forceCorrelation}
To obtain Weyl symbols that only depend on the system variables, 
one has to evaluate their mean value with respect to the reservoir degrees
of freedom. Generally, the mean value of a symbol $A(q,p)$
in phase space is expressed through
\begin{align} 
  \langle A \rangle =\int dq\, dp \, A(q,p)\, W(q,p).
\end{align} 
For a harmonic oscillator
in thermal equilibrium, the Wigner function 
takes the form \cite{Hillery1984121}
\begin{align} 
  W &= \frac{ \omega}{2\pi \langle E \rangle } 
   e^{-H(q,p)/\langle E \rangle }
\\ &=  \frac{ \omega}{2\pi \langle E \rangle } 
   e^{-\hbar\omega |\alpha|^2/\langle E \rangle }
\end{align} 
with $\langle E \rangle = 
\frac{ 1}{2} \hbar\omega \coth(\hbar\omega /(2k_BT))$ the mean energy and
$H(q,p)=p^2/(2m)+\frac{ 1}{2} m\omega^2 q^2$ the Weyl symbol
of the Hamiltonian.
This result can be derived using the thermal density
matrix $\rho = \exp[-\hat{H}/(k_B T)]/ \text{Tr}\exp[-\hat{H}/(k_B
T)]$ for the harmonic oscillator and
Mehler's formula (see Sec.~10.13 of Ref.~\cite{Oberhettinger}).
If we assume that the reservoir oscillators are thermalized,
one can easily evaluate that
$  \langle |\alpha_n|^2 \rangle =
   \langle E_n \rangle  / (\hbar \omega_n)$
and $\langle \alpha_n \rangle = \langle \alpha_n^2 \rangle =0$.
Eq.~(\ref{eq:Fdef}) then implies $\langle F(t) \rangle =0$ and
\begin{align} 
  \langle F(t) F(t')\rangle &= 
  \sum_n k_n^2l_n^2 \frac{ \langle E_n \rangle }{\hbar \omega_n} \cos(\omega_n(t-t')).
\end{align} 
For a hot reservoir, where $\langle E_n \rangle \gg \hbar \omega_n$,
one has $\langle E_n \rangle \approx k_B T$, from 
which Eq.~(\ref{eq:FFcorrelation}) follows.

\section{Solution for canonical observables without Markov approximation}\label{sec:LaplaceTreatment}
In this section we repeat the calculations of
Sec.~\ref{sec:canonicalSymbols} without making the 
Markov approximation (\ref{eq:MarkovApprox}).
This amounts to solving Eq.~(\ref{eq:freeMoyal})
with $-2\gamma  \delta \bar{A}/\delta F(t)$ replaced
by $K(t) \bar{A}/m$. By making the same ansatz as 
in Sec.~\ref{sec:canonicalSymbols} we obtain the same
equations and solutions for $\beta_2$ and $\bm{\beta}_3$,
but $\beta_1$ is determined by
\begin{align} 
  m\dot{\beta}_1 &= \beta_2 + K \bm{\beta}_3 .
\end{align} 
Using Eq.~(\ref{eq:KFaction}) this can be transformed into
\begin{align}
   m \dot{\beta}_1 &=  
    \beta_2(0) -
   \int_0^tdt'\, 
   \beta_1(t') C(t-t').
\label{eq:dglA1}\end{align} 
This equation can be solved using a Laplace transformation
$g(t)\rightarrow \tilde{g}(s)$,
\begin{align} 
  \tilde{\beta}_1(s) &= \frac{ 1}{s+\frac{ 1}{m} \tilde{C}(s)}
  \left (
  \beta_1(0) + \frac{ 1}{m s} \beta_2(0)
  \right ).
\end{align} 

As an example 
for a reservoir memory function we consider
\footnote{Eq.~(\ref{eq:defC}) suggests that 
 $\partial_t C(0)=0$. Although example (\ref{eq:fExample}) does not obey
this relation,
we have verified that in the limit $\Gamma \gg \gamma$
it yields the same result as more complicated
examples that do fulfill it.}
\begin{align} 
  C(t) &= m \gamma  \Gamma  e^{-\Gamma  t} 
\label{eq:fExample}\\
  \tilde{C}(s) &= \frac{m \gamma  \Gamma   }{ s+\Gamma},
\end{align} 
which fulfills
$\int_0^\infty C(t)dt=m\gamma$.
The parameter $\gamma$ should correspond to the decay rate associated
with dissipation, while $\tau = 1/\Gamma$ corresponds to the short
time scale on which $C(t)$ decays. 

If $s_i$ denotes the poles of $ \tilde{C}(s)$, then the inverse Laplace
transformation contains time-dependent exponentials $e^{t s_i}$
multiplied by constant factors. For $\Gamma \gg \gamma$, we 
can make a separate Taylor expansion 
of the poles in the exponent and the constant terms
around $\Gamma =\infty$.
 The poles $s_i$  are then approximately given by 
$-\gamma$ and  $-\Gamma$. After applying 
transformation (\ref{eq:AAbarTrafo}) we
obtain, to leading order in $1/\Gamma$,
$A_q(t)$ of Eq.~(\ref{eq:symq}) as well as 
\begin{align} 
  A_p &=    e^{-\gamma t}p  
  -m \gamma q (e^{-\gamma t} -e^{-\Gamma t})
  + \int_0^t dt' \,  e^{-\gamma t'}F(t-t').
\end{align}  
For times $t\gg \tau$ this solution
agrees with the Markovian result (\ref{eq:symp}). 
It only differs by the short-living term proportional to 
$e^{-\Gamma  t}$, which ensures that the correct initial
conditions are fulfilled.

\end{appendix}

\bibliographystyle{apsrev4-1}
\bibliography{/Users/pmarzlin/Documents/literatur/kpmJabRef.bib}

\begin{thebibliography}{31}%
\makeatletter
\providecommand \@ifxundefined [1]{%
 \@ifx{#1\undefined}
}%
\providecommand \@ifnum [1]{%
 \ifnum #1\expandafter \@firstoftwo
 \else \expandafter \@secondoftwo
 \fi
}%
\providecommand \@ifx [1]{%
 \ifx #1\expandafter \@firstoftwo
 \else \expandafter \@secondoftwo
 \fi
}%
\providecommand \natexlab [1]{#1}%
\providecommand \enquote  [1]{``#1''}%
\providecommand \bibnamefont  [1]{#1}%
\providecommand \bibfnamefont [1]{#1}%
\providecommand \citenamefont [1]{#1}%
\providecommand \href@noop [0]{\@secondoftwo}%
\providecommand \href [0]{\begingroup \@sanitize@url \@href}%
\providecommand \@href[1]{\@@startlink{#1}\@@href}%
\providecommand \@@href[1]{\endgroup#1\@@endlink}%
\providecommand \@sanitize@url [0]{\catcode `\\12\catcode `\$12\catcode
  `\&12\catcode `\#12\catcode `\^12\catcode `\_12\catcode `\%12\relax}%
\providecommand \@@startlink[1]{}%
\providecommand \@@endlink[0]{}%
\providecommand \url  [0]{\begingroup\@sanitize@url \@url }%
\providecommand \@url [1]{\endgroup\@href {#1}{\urlprefix }}%
\providecommand \urlprefix  [0]{URL }%
\providecommand \Eprint [0]{\href }%
\providecommand \doibase [0]{http://dx.doi.org/}%
\providecommand \selectlanguage [0]{\@gobble}%
\providecommand \bibinfo  [0]{\@secondoftwo}%
\providecommand \bibfield  [0]{\@secondoftwo}%
\providecommand \translation [1]{[#1]}%
\providecommand \BibitemOpen [0]{}%
\providecommand \bibitemStop [0]{}%
\providecommand \bibitemNoStop [0]{.\EOS\space}%
\providecommand \EOS [0]{\spacefactor3000\relax}%
\providecommand \BibitemShut  [1]{\csname bibitem#1\endcsname}%
\let\auto@bib@innerbib\@empty
\bibitem [{\citenamefont {Wigner}(1932)}]{PhysRev.40.749}%
  \BibitemOpen
  \bibfield  {author} {\bibinfo {author} {\bibfnamefont {E.}~\bibnamefont
  {Wigner}},\ }\href {\doibase 10.1103/PhysRev.40.749} {\bibfield  {journal}
  {\bibinfo  {journal} {Phys. Rev.}\ }\textbf {\bibinfo {volume} {40}},\
  \bibinfo {pages} {749} (\bibinfo {year} {1932})}\BibitemShut {NoStop}%
\bibitem [{\citenamefont {Schleich}(2001)}]{SchleichPhaseSpace}%
  \BibitemOpen
  \bibfield  {author} {\bibinfo {author} {\bibfnamefont {W.~P.}\ \bibnamefont
  {Schleich}},\ }\href@noop {} {\emph {\bibinfo {title} {Quantum Optics in
  Phase Space}}}\ (\bibinfo  {publisher} {Wiley-VCH},\ \bibinfo {year}
  {2001})\BibitemShut {NoStop}%
\bibitem [{\citenamefont {Weyl}(1927)}]{Weyl1927}%
  \BibitemOpen
  \bibfield  {author} {\bibinfo {author} {\bibfnamefont {H.}~\bibnamefont
  {Weyl}},\ }\href {\doibase 10.1007/BF02055756} {\bibfield  {journal}
  {\bibinfo  {journal} {Zeitschr. Phys.}\ }\textbf {\bibinfo {volume} {46}},\
  \bibinfo {pages} {1} (\bibinfo {year} {1927})}\BibitemShut {NoStop}%
\bibitem [{\citenamefont {Case}(2008)}]{case:937}%
  \BibitemOpen
  \bibfield  {author} {\bibinfo {author} {\bibfnamefont {W.~B.}\ \bibnamefont
  {Case}},\ }\href {\doibase 10.1119/1.2957889} {\bibfield  {journal} {\bibinfo
   {journal} {Am. J. Phys.}\ }\textbf {\bibinfo {volume} {76}},\ \bibinfo
  {pages} {937} (\bibinfo {year} {2008})}\BibitemShut {NoStop}%
\bibitem [{\citenamefont {Moyal}(1949)}]{PSP:1593184}%
  \BibitemOpen
  \bibfield  {author} {\bibinfo {author} {\bibfnamefont {J.~E.}\ \bibnamefont
  {Moyal}},\ }\href {\doibase 10.1017/S0305004100000487} {\bibfield  {journal}
  {\bibinfo  {journal} {Math. Proc. Camb. Phil. Soc.}\ }\textbf {\bibinfo
  {volume} {45}},\ \bibinfo {pages} {99} (\bibinfo {year} {1949})}\BibitemShut
  {NoStop}%
\bibitem [{\citenamefont {Groenewold}(1946)}]{Groenewold1946405}%
  \BibitemOpen
  \bibfield  {author} {\bibinfo {author} {\bibfnamefont {H.}~\bibnamefont
  {Groenewold}},\ }\href {\doibase
  http://dx.doi.org/10.1016/S0031-8914(46)80059-4} {\bibfield  {journal}
  {\bibinfo  {journal} {Physica}\ }\textbf {\bibinfo {volume} {12}},\ \bibinfo
  {pages} {405 } (\bibinfo {year} {1946})}\BibitemShut {NoStop}%
\bibitem [{\citenamefont {Osborn}\ and\ \citenamefont
  {Molzahn}(1995)}]{Osborn199579}%
  \BibitemOpen
  \bibfield  {author} {\bibinfo {author} {\bibfnamefont {T.~A.}\ \bibnamefont
  {Osborn}}\ and\ \bibinfo {author} {\bibfnamefont {F.~H.}\ \bibnamefont
  {Molzahn}},\ }\href {\doibase DOI: 10.1006/aphy.1995.1057} {\bibfield
  {journal} {\bibinfo  {journal} {Ann. Phys.}\ }\textbf {\bibinfo {volume}
  {241}},\ \bibinfo {pages} {79 } (\bibinfo {year} {1995})}\BibitemShut
  {NoStop}%
\bibitem [{\citenamefont {Ozorio{\ }de{\ }Almeida}(1998)}]{deAlmeida1998265}%
  \BibitemOpen
  \bibfield  {author} {\bibinfo {author} {\bibfnamefont {A.~M.}\ \bibnamefont
  {Ozorio{\ }de{\ }Almeida}},\ }\href {\doibase 10.1016/S0370-1573(97)00070-7}
  {\bibfield  {journal} {\bibinfo  {journal} {Phys. Rep.}\ }\textbf {\bibinfo
  {volume} {295}},\ \bibinfo {pages} {265 } (\bibinfo {year}
  {1998})}\BibitemShut {NoStop}%
\bibitem [{\citenamefont {Stratonovich}(1957)}]{STR57}%
  \BibitemOpen
  \bibfield  {author} {\bibinfo {author} {\bibfnamefont {R.~L.}\ \bibnamefont
  {Stratonovich}},\ }\href@noop {} {\bibfield  {journal} {\bibinfo  {journal}
  {Sov. Phys. JETP}\ }\textbf {\bibinfo {volume} {4}},\ \bibinfo {pages} {891}
  (\bibinfo {year} {1957})}\BibitemShut {NoStop}%
\bibitem [{\citenamefont {Berezin}(1974)}]{FAB72}%
  \BibitemOpen
  \bibfield  {author} {\bibinfo {author} {\bibfnamefont {F.~A.}\ \bibnamefont
  {Berezin}},\ }\href@noop {} {\bibfield  {journal} {\bibinfo  {journal} {Math.
  USSR-Izv.}\ }\textbf {\bibinfo {volume} {8}},\ \bibinfo {pages} {1109}
  (\bibinfo {year} {1974})}\BibitemShut {NoStop}%
\bibitem [{\citenamefont {Grossmann}(1976)}]{Gro76}%
  \BibitemOpen
  \bibfield  {author} {\bibinfo {author} {\bibfnamefont {A.}~\bibnamefont
  {Grossmann}},\ }\href {http://projecteuclid.org/euclid.cmp/1103899886}
  {\bibfield  {journal} {\bibinfo  {journal} {Commun. Math. Phys.}\ }\textbf
  {\bibinfo {volume} {48}},\ \bibinfo {pages} {191} (\bibinfo {year}
  {1976})}\BibitemShut {NoStop}%
\bibitem [{\citenamefont {Royer}(1977)}]{Roy77}%
  \BibitemOpen
  \bibfield  {author} {\bibinfo {author} {\bibfnamefont {A.}~\bibnamefont
  {Royer}},\ }\href {\doibase 10.1103/PhysRevA.15.449} {\bibfield  {journal}
  {\bibinfo  {journal} {Phys. Rev. A}\ }\textbf {\bibinfo {volume} {15}},\
  \bibinfo {pages} {449} (\bibinfo {year} {1977})}\BibitemShut {NoStop}%
\bibitem [{\citenamefont {Osborn}\ and\ \citenamefont
  {Marzlin}(2009)}]{jpa42-415302}%
  \BibitemOpen
  \bibfield  {author} {\bibinfo {author} {\bibfnamefont {T.~A.}\ \bibnamefont
  {Osborn}}\ and\ \bibinfo {author} {\bibfnamefont {K.-P.}\ \bibnamefont
  {Marzlin}},\ }\href {\doibase 10.1088/1751-8113/42/41/415302} {\bibfield
  {journal} {\bibinfo  {journal} {J. Phys. A}\ }\textbf {\bibinfo {volume}
  {42}},\ \bibinfo {pages} {415302} (\bibinfo {year} {2009})}\BibitemShut
  {NoStop}%
\bibitem [{\citenamefont {Gardiner}\ and\ \citenamefont
  {Zoller}(2004)}]{GardinerZollerQuantumNoise}%
  \BibitemOpen
  \bibfield  {author} {\bibinfo {author} {\bibfnamefont {C.~W.}\ \bibnamefont
  {Gardiner}}\ and\ \bibinfo {author} {\bibfnamefont {P.}~\bibnamefont
  {Zoller}},\ }\href
  {http://books.google.ca/books?id=a_xsT8oGhdgC&lpg=PP1&ots=kXx1vVbWu9&dq=gardiner%20zoller&pg=PR4#v=onepage&q&f=false}
  {\emph {\bibinfo {title} {Quantum Noise}}}\ (\bibinfo  {publisher} {Springer,
  Berlin},\ \bibinfo {year} {2004})\BibitemShut {NoStop}%
\bibitem [{\citenamefont {Breuer}\ and\ \citenamefont
  {Petruccione}(2007)}]{BreuerPettrucione}%
  \BibitemOpen
  \bibfield  {author} {\bibinfo {author} {\bibfnamefont {H.~P.}\ \bibnamefont
  {Breuer}}\ and\ \bibinfo {author} {\bibfnamefont {F.}~\bibnamefont
  {Petruccione}},\ }\href@noop {} {\emph {\bibinfo {title} {The Theory of Open
  Quantum Systems}}}\ (\bibinfo  {publisher} {Oxford University Press},\
  \bibinfo {year} {2007})\BibitemShut {NoStop}%
\bibitem [{\citenamefont {Busch}\ \emph {et~al.}(1996)\citenamefont {Busch},
  \citenamefont {Lahti},\ and\ \citenamefont {Mittelstaedt}}]{BuschLahtiMitt}%
  \BibitemOpen
  \bibfield  {author} {\bibinfo {author} {\bibfnamefont {P.}~\bibnamefont
  {Busch}}, \bibinfo {author} {\bibfnamefont {P.}~\bibnamefont {Lahti}}, \ and\
  \bibinfo {author} {\bibfnamefont {P.}~\bibnamefont {Mittelstaedt}},\ }\href
  {http://link.springer.com/book/10.1007/978-3-540-37205-9} {\emph {\bibinfo
  {title} {The quantum theory of measurement}}}\ (\bibinfo  {publisher}
  {Springer},\ \bibinfo {year} {1996})\BibitemShut {NoStop}%
\bibitem [{\citenamefont {Lindblad}(1976)}]{linblad:cmp76}%
  \BibitemOpen
  \bibfield  {author} {\bibinfo {author} {\bibfnamefont {G.}~\bibnamefont
  {Lindblad}},\ }\href {\doibase 10.1007/BF01608499} {\bibfield  {journal}
  {\bibinfo  {journal} {Comm. Math. Phys.}\ }\textbf {\bibinfo {volume} {48}},\
  \bibinfo {pages} {119} (\bibinfo {year} {1976})}\BibitemShut {NoStop}%
\bibitem [{\citenamefont {Di\'osi}\ and\ \citenamefont
  {Kiefer}(2002)}]{0305-4470-35-11-312}%
  \BibitemOpen
  \bibfield  {author} {\bibinfo {author} {\bibfnamefont {L.}~\bibnamefont
  {Di\'osi}}\ and\ \bibinfo {author} {\bibfnamefont {C.}~\bibnamefont
  {Kiefer}},\ }\href {http://stacks.iop.org/0305-4470/35/i=11/a=312} {\bibfield
   {journal} {\bibinfo  {journal} {J. Phys. A}\ }\textbf {\bibinfo {volume}
  {35}},\ \bibinfo {pages} {2675} (\bibinfo {year} {2002})}\BibitemShut
  {NoStop}%
\bibitem [{\citenamefont {Brodier}\ and\ \citenamefont {Ozorio{\ }de{\
  }Almeida}(2004)}]{PhysRevE.69.016204}%
  \BibitemOpen
  \bibfield  {author} {\bibinfo {author} {\bibfnamefont {O.}~\bibnamefont
  {Brodier}}\ and\ \bibinfo {author} {\bibfnamefont {A.~M.}\ \bibnamefont
  {Ozorio{\ }de{\ }Almeida}},\ }\href {\doibase 10.1103/PhysRevE.69.016204}
  {\bibfield  {journal} {\bibinfo  {journal} {Phys. Rev. E}\ }\textbf {\bibinfo
  {volume} {69}},\ \bibinfo {pages} {016204} (\bibinfo {year}
  {2004})}\BibitemShut {NoStop}%
\bibitem [{\citenamefont {Zurek}(2003)}]{RevModPhys.75.715}%
  \BibitemOpen
  \bibfield  {author} {\bibinfo {author} {\bibfnamefont {W.~H.}\ \bibnamefont
  {Zurek}},\ }\href {\doibase 10.1103/RevModPhys.75.715} {\bibfield  {journal}
  {\bibinfo  {journal} {Rev. Mod. Phys.}\ }\textbf {\bibinfo {volume} {75}},\
  \bibinfo {pages} {715} (\bibinfo {year} {2003})}\BibitemShut {NoStop}%
\bibitem [{\citenamefont {Agarwal}(1971)}]{PhysRevA.4.739}%
  \BibitemOpen
  \bibfield  {author} {\bibinfo {author} {\bibfnamefont {G.}~\bibnamefont
  {Agarwal}},\ }\href {\doibase 10.1103/PhysRevA.4.739} {\bibfield  {journal}
  {\bibinfo  {journal} {Phys. Rev. A}\ }\textbf {\bibinfo {volume} {4}},\
  \bibinfo {pages} {739} (\bibinfo {year} {1971})}\BibitemShut {NoStop}%
\bibitem [{\citenamefont {Dalibard}\ and\ \citenamefont
  {Cohen-Tannoudji}(1985)}]{0022-3700-18-8-019}%
  \BibitemOpen
  \bibfield  {author} {\bibinfo {author} {\bibfnamefont {J.}~\bibnamefont
  {Dalibard}}\ and\ \bibinfo {author} {\bibfnamefont {C.}~\bibnamefont
  {Cohen-Tannoudji}},\ }\href {\doibase 10.1088/0022-3700/18/8/019} {\bibfield
  {journal} {\bibinfo  {journal} {J. Phys. B}\ }\textbf {\bibinfo {volume}
  {18}},\ \bibinfo {pages} {1661} (\bibinfo {year} {1985})}\BibitemShut
  {NoStop}%
\bibitem [{\citenamefont {Ozorio{\ }de{\ }Almeida}(2008)}]{quant-ph/0612029}%
  \BibitemOpen
  \bibfield  {author} {\bibinfo {author} {\bibfnamefont {A.~M.}\ \bibnamefont
  {Ozorio{\ }de{\ }Almeida}},\ }\href {http://arxiv.org/abs/quant-ph/0612029}
  {\bibfield  {journal} {\bibinfo  {journal} {in Entanglement and Decoherence,
  A. Buchleitner, C. Viviescas, M. Tiersch (Editors) Springer
  (quant-ph/0612029)}\ } (\bibinfo {year} {2008})}\BibitemShut {NoStop}%
\bibitem [{\citenamefont {Ford}\ \emph {et~al.}(1965)\citenamefont {Ford},
  \citenamefont {Kac},\ and\ \citenamefont {Mazur}}]{FordKacMazur}%
  \BibitemOpen
  \bibfield  {author} {\bibinfo {author} {\bibfnamefont {G.~W.}\ \bibnamefont
  {Ford}}, \bibinfo {author} {\bibfnamefont {M.}~\bibnamefont {Kac}}, \ and\
  \bibinfo {author} {\bibfnamefont {P.}~\bibnamefont {Mazur}},\ }\href
  {\doibase http://dx.doi.org/10.1063/1.1704304} {\bibfield  {journal}
  {\bibinfo  {journal} {J. Math. Phys.}\ }\textbf {\bibinfo {volume} {6}},\
  \bibinfo {pages} {504} (\bibinfo {year} {1965})}\BibitemShut {NoStop}%
\bibitem [{\citenamefont {Ford}\ and\ \citenamefont {Kac}(1987)}]{FordKac1987}%
  \BibitemOpen
  \bibfield  {author} {\bibinfo {author} {\bibfnamefont {G.}~\bibnamefont
  {Ford}}\ and\ \bibinfo {author} {\bibfnamefont {M.}~\bibnamefont {Kac}},\
  }\href {\doibase 10.1007/BF01011142} {\bibfield  {journal} {\bibinfo
  {journal} {J. Stat. Phys.}\ }\textbf {\bibinfo {volume} {46}},\ \bibinfo
  {pages} {803} (\bibinfo {year} {1987})}\BibitemShut {NoStop}%
\bibitem [{\citenamefont {Marzlin}\ \emph {et~al.}(2010)\citenamefont
  {Marzlin}, \citenamefont {Wang}, \citenamefont {Moiseev},\ and\ \citenamefont
  {Sanders}}]{JOSA-B27-A36}%
  \BibitemOpen
  \bibfield  {author} {\bibinfo {author} {\bibfnamefont {K.-P.}\ \bibnamefont
  {Marzlin}}, \bibinfo {author} {\bibfnamefont {Z.-B.}\ \bibnamefont {Wang}},
  \bibinfo {author} {\bibfnamefont {S.~A.}\ \bibnamefont {Moiseev}}, \ and\
  \bibinfo {author} {\bibfnamefont {B.~C.}\ \bibnamefont {Sanders}},\ }\href
  {\doibase 10.1364/JOSAB.27.000A36} {\bibfield  {journal} {\bibinfo  {journal}
  {J. Opt. Soc. Am. B}\ }\textbf {\bibinfo {volume} {27}},\ \bibinfo {pages}
  {A36} (\bibinfo {year} {2010})}\BibitemShut {NoStop}%
\bibitem [{\citenamefont {Meng}\ \emph {et~al.}(2009)\citenamefont {Meng},
  \citenamefont {Wang},\ and\ \citenamefont {Liang}}]{Meng20092027}%
  \BibitemOpen
  \bibfield  {author} {\bibinfo {author} {\bibfnamefont {X.-G.}\ \bibnamefont
  {Meng}}, \bibinfo {author} {\bibfnamefont {J.-S.}\ \bibnamefont {Wang}}, \
  and\ \bibinfo {author} {\bibfnamefont {B.-L.}\ \bibnamefont {Liang}},\ }\href
  {\doibase http://dx.doi.org/10.1016/j.ssc.2009.08.028} {\bibfield  {journal}
  {\bibinfo  {journal} {Sol. State Comm.}\ }\textbf {\bibinfo {volume} {149}},\
  \bibinfo {pages} {2027 } (\bibinfo {year} {2009})}\BibitemShut {NoStop}%
\bibitem [{\citenamefont {Boivin}\ \emph {et~al.}(1994)\citenamefont {Boivin},
  \citenamefont {K\"artner},\ and\ \citenamefont {Haus}}]{PRL73:240}%
  \BibitemOpen
  \bibfield  {author} {\bibinfo {author} {\bibfnamefont {L.}~\bibnamefont
  {Boivin}}, \bibinfo {author} {\bibfnamefont {F.~X.}\ \bibnamefont
  {K\"artner}}, \ and\ \bibinfo {author} {\bibfnamefont {H.~A.}\ \bibnamefont
  {Haus}},\ }\href {http://link.aps.org/abstract/PRL/v73/p240} {\bibfield
  {journal} {\bibinfo  {journal} {Phys. Rev. Lett.}\ }\textbf {\bibinfo
  {volume} {73}},\ \bibinfo {pages} {240} (\bibinfo {year} {1994})}\BibitemShut
  {NoStop}%
\bibitem [{\citenamefont {Shapiro}(2006)}]{shapiro:062305}%
  \BibitemOpen
  \bibfield  {author} {\bibinfo {author} {\bibfnamefont {J.~H.}\ \bibnamefont
  {Shapiro}},\ }\href {\doibase 10.1103/PhysRevA.73.062305} {\bibfield
  {journal} {\bibinfo  {journal} {Phys. Rev. A}\ }\textbf {\bibinfo {volume}
  {73}},\ \bibinfo {eid} {062305} (\bibinfo {year} {2006})}\BibitemShut
  {NoStop}%
\bibitem [{\citenamefont {Hillery}\ \emph {et~al.}(1984)\citenamefont
  {Hillery}, \citenamefont {O'Connell}, \citenamefont {Scully},\ and\
  \citenamefont {Wigner}}]{Hillery1984121}%
  \BibitemOpen
  \bibfield  {author} {\bibinfo {author} {\bibfnamefont {M.}~\bibnamefont
  {Hillery}}, \bibinfo {author} {\bibfnamefont {R.}~\bibnamefont {O'Connell}},
  \bibinfo {author} {\bibfnamefont {M.}~\bibnamefont {Scully}}, \ and\ \bibinfo
  {author} {\bibfnamefont {E.}~\bibnamefont {Wigner}},\ }\href {\doibase
  http://dx.doi.org/10.1016/0370-1573(84)90160-1} {\bibfield  {journal}
  {\bibinfo  {journal} {Phys. Rep.}\ }\textbf {\bibinfo {volume} {106}},\
  \bibinfo {pages} {121 } (\bibinfo {year} {1984})}\BibitemShut {NoStop}%
\bibitem [{\citenamefont {Magnus}\ \emph {et~al.}(1953)\citenamefont {Magnus},
  \citenamefont {Oberhettinger},\ and\ \citenamefont
  {Tricomi}}]{Oberhettinger}%
  \BibitemOpen
  \bibfield  {author} {\bibinfo {author} {\bibfnamefont {W.}~\bibnamefont
  {Magnus}}, \bibinfo {author} {\bibfnamefont {F.}~\bibnamefont
  {Oberhettinger}}, \ and\ \bibinfo {author} {\bibfnamefont {F.~G.}\
  \bibnamefont {Tricomi}},\ }\href@noop {} {\emph {\bibinfo {title} {Higher
  transcendental functions}}},\ edited by\ \bibinfo {editor} {\bibfnamefont
  {A.}~\bibnamefont {Erd\'elyi}}\ (\bibinfo  {publisher} {McGraw Hill},\
  \bibinfo {year} {1953})\BibitemShut {NoStop}%
\end{thebibliography}%
\end{document}